\documentclass[pra,twocolumn]{revtex4}
\usepackage{graphicx}

\begin{document}
\title{Generation of atom-photon entangled states in atomic Bose-Einstein condensate
via electromagnetically induced transparency}
\author{Le-Man Kuang$^{1,2}$ and Lan Zhou$^{1}$}
\address{$^{1}$Department of Physics, Hunan Normal University, Changsha 410081, People's Republic of China\\
$^{2}$The Abdus Salam International Centre for Theoretical Physics, Strada Costiera 11, Trieste 34014, Italy}
\begin{abstract}
 In this paper, we present a method to generate continuous-variable-type entangled states
 between photons and atoms in atomic Bose-Einstein condensate (BEC). The proposed method involves an
 atomic BEC with three internal states, a weak quantized probe laser and a strong classical coupling laser,
 which form a three-level $\Lambda$-shaped BEC system. We consider a situation where the BEC is
 in electromagnetically induced transparency (EIT) with the coupling laser being much stronger than
 the probe laser. In this case, the upper and intermediate levels are unpopulated, so that their adiabatic
 elimination enables an effective two-mode model involving only the atomic field at the lowest internal
 level and the quantized probe laser field. Atom-photon quantum entanglement is created through laser-atom
 and inter-atomic interactions, and two-photon detuning. We show how to generate atom-photon entangled
 coherent states and entangled states between photon (atom) coherent states and atom-(photon-) macroscopic quantum superposition (MQS) states,
 and between photon-MQS  and atom-MQS states.

\noindent PACS number(s): 03.65.Ta, 03.65.Ud, 03.75.Fi
\end{abstract}

\maketitle

\section{Introduction}

For decades, quantum entanglement has been the focus of much work in
the foundations of quantum mechanics, being particularly associated with quantum
nonseparability, the violation of Bell's inequalities, and the so-called
Einstein-Pololsky-Rosen (EPR) paradox. Beyond this fundamental aspect,
creating and manipulating of entangled states are essential for quantum information applications.
Among these applications are quantum computation \cite{div,sho}, quantum
teleportation \cite{ben1,bou}, quantum dense coding \cite{ben2}, quantum
cryptography \cite{ben3}, and quantum positioning and clock synchronization
\cite{gio}. Hence, quantum entanglement has been viewed as an essential
resource for quantum information processing.

Recently much attention has been paid to continuous variable quantum information processing in which continuous-variable-type entangled pure states play a key role. For instance, two-state entangled coherent states are used to realize efficient quantum computation \cite{jeo,mun} and quantum teleportation \cite{enk}. Two-mode squeezed vacuum states have been applied to quantum dense coding \cite{ban}.  Especially, continuous variable teleportation \cite{bra} has been experimentally demonstrated for coherent states of a light field \cite{fur} by using entangled two-mode squeezed vacuum. It is also has been shown that a two-state entangled squeezed vacuum state can be used to realize quantum teleportation of an arbitrary coherent superposition state of two equal-amplitude and opposite-phase squeezed vacuum states \cite{cai}. Continuous-variable-type entangled states including squeezed states and coherent states have also been widely applied to and quantum cryptography \cite{ral,hil,rei, gro,got,cer}. Therefore, it is an interesting topic to create continuous-variable-type entangled pure states.

On the other hand, it is well known that atoms and photons can be viewed as carriers of storing and transmitting quantum information. Atoms are suited for storing quantum information at a fixed location, and they are sources of local entanglement for quantum information processing while photons are natural sources for communications of quantum information since they can transverse over long distances.
One topic of particular interest in this context is the possibility to generate atom-photon entangled states where one member is readily used to couple a distant system while the other is stored by the local sender.
In this aspect, some progress has been made, and possible schemes have been proposed in recent years. For instance,
Moore and Meystre \cite{moo} proposed a scheme to generate atom-photon pairs via off-resonant light scattering from an atomic Bose-Einstein condensate (BEC). Deb and Agarwal \cite{deb} showed that it is possible to entangle three different many-particle states by Bragg spectroscopy with nonclassical light in an atomic BEC. Gasenzer and co-workers \cite{gas} investigated quantum entanglement characterized by the relative number squeezing between photons and atoms coupled out from an atomic BEC. Optical control over the BEC quantum statistics and atom-photon quantum correlation in an atomic BEC \cite{pra,zen,moor,kua} have been widely studied.

In this paper, we describe a method to produce atom-photon-entangled states of
continuous-variable-type pure states in an atomic BEC by electromagnetically induced transparency (EIT) \cite{har}. EIT is a kind of quantum interference effects \cite{scu}. It arises in three-level (or multilevel) systems and consists of the cancellation of the absorption on one transition induced by simultaneous coherent driving of another transition. The phenomenon can be understood as a destructive interference of the two pathways to the excited level and has been used to demonstrate ultraslow light propagation \cite{hau,ino,kas,bud,tur} and light storage \cite{liu,phi} in many systems including atomic BECs \cite{hau,ino}.
The proposed system in present paper consists of an atomic BEC with three internal states, one weak probe laser, and one relatively strong coupling laser with appropriate frequencies. The probe laser field is quantized while the coupling laser field is treated classically. They form a three-level lambda configuration.
When the BEC is in EIT with the coupling laser being much stronger than the probe laser, the upper and intermediate levels are unpopulated due to quantum interference. Adiabatic elimination of the two unpopulated levels enables an effective two-mode model involving only atoms in the lowest internal state and photons in the probe laser.
We show that it is possible to generate atom-photon continuous-variable-type entangled pure states in the atomic BEC.
This paper is organized as follows. In Sec. II, we present the physical system under our consideration, review the reduction of the problem from four to two modes, followed by an analytic solution of the model, and then discuss the theoretical mechanism to create quantum entanglement in the proposed scheme. In Sec. III, we focus on a dynamical approach to creating a variety of atom-photon entangled pure states. Two-state, three-state, and four-state atom-photon continuous-variable-typed entangled pure states are generated explicitly.
We shall conclude our paper with discussions and remarks in the last section.

\section{Physical model and solution}

We consider a cloud of BEC atoms which have three internal states labeled by $|1\rangle$,  $|2\rangle$, and  $|3\rangle$ with
energies $E_1$,   $E_2$, and $E_3$, respectively.
The two lower states   $|1\rangle$ and  $|3\rangle$ are metastable states in each of which the atoms can live for a long time. They are Raman coupled to
the  upper state  $|2\rangle$ via, respectively, a quantized probe laser field and a classical coupling laser field of frequencies $\omega_1$ and $\omega_2$  in the Lambda configuration. The interaction scheme is shown in Fig. 1.
The atoms in these internal states are subject to isotropic harmonic trapping potentials $V_i({\bf r})$ for $i=1,2,3$, respectively. Furthermore, the atoms in the BEC interact with each other via elastic two-body collisions with the $\delta$-function potentials $V_{ij}({\bf r}-{\bf r}')=U_{ij}\delta ({\bf r}-{\bf r}')$, where $U_{ij}=4\pi\hbar^2a_{ij}/m$ with $m$ and  $a_{ij}$, respectively, being the atomic mass and  the $s$-wave scattering length between atoms in states $i$ and $j$. A good experimental candidate of this system is the sodium atom condensate for which there exist appropriate atomic internal levels and external laser fields to form the Lambda configuration which is needed for reaching EIT under our consideration. The sodium atom condensate in EIT has been used to demonstrate ultraslow light propagation \cite{hau} and amplification of light and atoms \cite{ino} in atomic BECs.

\begin{figure}[htb]
\begin{center}
\includegraphics[width=7cm]{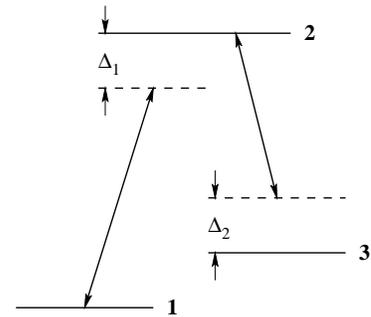}
\end{center}
\vskip 0.2cm
\caption{Three-level $\Lambda$-shaped atoms coupled to a quantized probe laser and a classical coupling laser with the detunings $\Delta_1$ and $\Delta_2$, respectively.}
\label{fig1}
\end{figure}

The second quantized Hamiltonian to describe the system at zero temperature is given by
\begin{equation}
\label{1}
\hat{H}=\hat{H}_{p}+\hat{H}_{a}+\hat{H}_{a-l}+\hat{H}_{c},
\end{equation}
where $\hat{H}_{p}$ and $\hat{H}_{a}$ gives the free evolution of the probe laser field and the atomic fields respectively, $\hat{H}_{a-l}$ describes the dipole interactions between the atomic fields and the probe and coupling fields, and $\hat{H}_{c}$ represents inter-atom two-body collisional interactions.

The free atomic Hamiltonian is given by
\begin{equation}
\label{2} \hat{H}_{a}=\sum^3_{i=1}\int d{\bf x}
\hat{\psi}^{\dagger}_i({\bf x}) \left [-\frac{\hbar^2}{2m}\nabla^2
+V_i({\bf x})+E_i\right ] \hat{\psi}_i({\bf x}),
\end{equation}
where $E_i$ are internal energies for the three internal states, $\hat{\psi}_i({\bf x})$ and $\hat{\psi}^{\dagger}_i({\bf x})$ are
the boson annihilation  and creation operators for the $|i\rangle$-state atoms at position ${\bf x}$, respectively, they  satisfy   the standard boson commutation relation $[\hat{\psi}_i({\bf x}), \hat{\psi}^{\dagger}_j({\bf x}')]=\delta_{ij}\delta({\bf x}-{\bf x}')$ and
$[\hat{\psi}_i({\bf x}), \hat{\psi}_j({\bf x}')]
=0=[\hat{\psi}^{\dagger}_i({\bf x}), \hat{\psi}^{\dagger}_j({\bf x}')]$.

The free evolution of the probe laser field is governed by the Hamiltonian
\begin{equation}
\label{3} \hat{H}_{p}=\hbar\omega_1\hat{a}^{\dagger}_1\hat{a}_1,
\end{equation}
where $\omega_1$ is the frequency of the probe laser, and $ \hat{a}^{\dagger}_1$ and $\hat{a}_1$ are the photon creation and annihilation operators for the probe laser field, satisfying the boson communication relation $[\hat{a}_1, \hat{a}^{\dagger}_1]=1$.

The atom-laser interactions in the dipole approximation can be described by the following Hamiltonian
\begin{eqnarray}
\label{4}
\hat{H}_{a-l}&=&-\hbar \int d{\bf x}  \left [g\hat{a}_1\hat{\psi}^{\dagger}_2({\bf x})\hat{\psi}_1({\bf x})e^{i({\bf k_1}\cdot {\bf x}-\omega_1t)} \right. \nonumber \\
& &\left.+\frac{1}{2}\Omega\hat{\psi}^{\dagger}_2({\bf x})
\hat{\psi}_3({\bf x})e^{i({\bf k_2}\cdot {\bf x}-\omega_2t)}+H.c.\right ],
\end{eqnarray}
where the two  dipole coupling constants are defined by
$g=\mu_{21}{\cal E}_1/\hbar$ and $\Omega=\mu_{23}E_2/\hbar$ with
$\mu_{ij}$ denoting a transition dipole-matrix element between states $|i\rangle$ and $|j\rangle$, ${\cal E}_1=\sqrt{\hbar\omega_1/2\epsilon_0V}$
being the electric field per photon for the quantized probe light of frequency $\omega_1$ in a mode volume $V$, and $E_2$ being the amplitude of the electric field for the classical coupling light of frequency $\omega_2$, ${\bf k}_1$ and ${\bf k}_2$ are wave vectors of correspondent laser fields.

The collisional Hamiltonian is taken to be the following form
\begin{eqnarray}
\label{5}
\hat{H}_{c}&=&\frac{2\pi\hbar^2}{m}
\int d{\bf x}\left [\sum^3_{i=1}a^{sc}_i\hat{\psi}^{\dagger}_i({\bf x})
\hat{\psi}^{\dagger}_i({\bf x})\hat{\psi}_i({\bf x})\hat{\psi}_i({\bf x})
\right. \nonumber \\
& &\left.+\sum_{i\neq j}2a^{sc}_{ij} \hat{\psi}^{\dagger}_i({\bf x})
\hat{\psi}^{\dagger}_j({\bf x})\hat{\psi}_i({\bf x})\hat{\psi}_j({\bf x})
\right ],
\end{eqnarray}
where $a^{sc}_i$ is  the $s$-wave scattering length of condensed
atoms  in the internal state $|i\rangle$ and $a^{sc}_{ij}$ that
between condensed atoms in the internal states $|i\rangle$ and
$|j\rangle$.

For a weakly interacting BEC at zero temperature there are no
thermally  excited atoms and the quantum depletion is negligible,
the motional state is frozen to be approximately the ground state.
One may neglect all modes except for the condensate mode and
approximately factorize the atomic field operators as
$\hat{\psi}_i({\bf x})\approx \hat{b}_i\phi_i({\bf x})$ where
$\phi_i({\bf x})$  is a normalized wavefunction for the atoms in
the BEC in the internal state $|i\rangle$,  which is given by the
ground state of the following Schr\"{o}dinger equation
\begin{equation}
\label{5'} \left [-\frac{\hbar^2}{2m}\nabla^2 +V_i({\bf
x})+E_i\right ] {\phi}_i({\bf x})=\hbar\nu_i{\phi}_i({\bf x}),
\end{equation}
where $\hbar\nu_i$ is the energy of the mode $i$, and $\nu_i$
denotes the frequency to the free evolution of the condensate in
the internal state $|i\rangle$.

Substituting the single-mode expansions of the atomic field
operators into Eqs. (\ref{2}), (\ref{4}), and   (\ref{5}), we
arrive at  the following four-mode approximate Hamiltonian
\begin{eqnarray}
\label{6}
\hat{H}&=&\hbar\omega_1\hat{a}^{\dagger}_1\hat{a}_1 + \hbar\sum^3_{i=1}\nu_i\hat{b}^{\dagger}_i\hat{b}_i \nonumber \\
&& -\hbar\left [g_1\hat{a}_1\hat{b}^{\dagger}_2\hat{b}_1e^{-i\omega_1t}
+g_2\hat{b}^{\dagger}_2\hat{b}_3e^{-i\omega_2t}+H.c. \right ]
\nonumber \\
&&+ \hbar\sum^3_{i=1}\lambda_i\hat{b}^{\dagger 2}_i\hat{b}^2_i
+ \hbar\sum_{i\neq j}\lambda_{ij}\hat{b}^{\dagger}_i\hat{b}_i
\hat{b}^{\dagger}_j\hat{b}_j,
\end{eqnarray}
where one mode corresponds to the probe laser field, the other
three to atomic fields in the three internal state.  Here  $g_1$
and $g_2$ denote the laser-atom dipole interactions, respectively,
defined by
\begin{eqnarray}
\label{6'}
  g_1&=&g\int d{\bf x} \phi^*_{2}({\bf x})\phi_{1}({\bf x})e^{i{\bf k}_1.{\bf
  x}},\\
\label{8'}
 g_2&=&\frac{1}{2}\Omega\int d{\bf x} \phi^*_{2}({\bf x})\phi_{3}({\bf x})e^{i{\bf k}_2.{\bf x}},
\end{eqnarray}
and $\lambda_i$ and $\lambda_{ij}$ ($i, j=1, 2, 3$)describe
inter-atomic interactions given by
\begin{eqnarray}
\label{8}
\lambda_i&=& \frac{2\pi\hbar^2a^{sc}_i}{m}\int d{\bf x}|\phi_{i}({\bf x})|^4,\\
\label{9}
  \lambda_{ij}&=&\frac{4\pi\hbar^2a^{sc}_{ij}}{m}\int d{\bf x}
|\phi_{i}({\bf x})|^2|\phi_{j}({\bf x})|^2,
 \hspace{0.3cm}(i\neq j).
\end{eqnarray}

The Hamiltonian  (\ref{6}) can be decomposed to the sum of an interaction Hamiltonian and a free evolution Hamiltonian defined by
\begin{eqnarray}
\label{7}
H_0&=&\hbar\omega_1\hat{a}^{\dagger}_1\hat{a}_1+\hbar\nu_1\sum^3_{i=1}\hat{b}^{\dagger}_i\hat{b}_i \nonumber \\
& &+\hbar(\omega_1-\omega_2)\hat{b}^{\dagger}_3\hat{b}_3 +\hbar\omega_1\hat{b}^{\dagger}_2\hat{b}_2.
\end{eqnarray}

Going over to an interaction picture with respect to $H_0$,  we transfer the time-dependent Hamiltonian (\ref{6}) to the following
 time-independent Hamiltonian
\begin{eqnarray}
\label{8}
 \hat{H}_I &=&\hbar(\Delta_1-\Delta_2)\hat{b}^{\dagger}_3\hat{b}_3+\Delta_1\hat{b}^{\dagger}_2\hat{b}_2
\nonumber \\
& &-\hbar[g_1\hat{a}_1\hat{b}^{\dagger}_2\hat{b}_1+ g_2\hat{b}^{\dagger}_2\hat{b}_3+H.c.]
\nonumber \\
& & + \hbar\sum^3_{i=1}\lambda_i\hat{b}^{\dagger 2}_i\hat{b}^2_i
+ \hbar\sum_{i\neq j}\lambda_{ij}\hat{b}^{\dagger}_i\hat{b}_i
\hat{b}^{\dagger}_j\hat{b}_j,
\end{eqnarray}
where $\Delta_1=\nu_2-\nu_1-\omega_1$ and $\Delta_2=\nu_2-\nu_3-\omega_2$
are the detunings of the two laser beams, respectively.

We consider the situation of the ideal EIT which is attained only
when the system is at the two-photon resonance with the two-photon
detuning  $\Delta =\Delta_1=\Delta_2$. Initially the lasers are
outside the BEC medium in which all atoms are in their ground
state, i.e., the internal state $|1\rangle$. The condensed atoms
are generally in a superposition state of the state $|1\rangle$
and the state $|3\rangle$ when they are in EIT. However, when the
coupling laser is much stronger than the probe laser, atomic
population at the intermediate level approaches zero while the
upper level is unpopulated \cite{scul}. Hence, under the condition
of $(g_{1}/g_{2})^{2}<<1$, the terms which involve
$\hat{b}^{\dagger}_2\hat{b}_2$ and $\hat{b}^{\dagger}_3\hat{b}_3$
in the Hamiltonian (\ref{8}) may be neglected, and from the
Hamiltonian (\ref{8}) we can adiabatically eliminate the atomic
field operators in the internal states $|2\rangle$ and
$|3\rangle$. Firstly, the atomic operators in the internal state
$|2\rangle$  $\hat{b}_2$ and $\hat{b}^{\dagger}_2$ can be
adiabatically eliminated with the replacements:
$\hat{b}_2=(g_1\hat{a}_1\hat{b}_1 + g_2\hat{b}_3)/\Delta $ and
$\hat{b}^{\dagger}_2=(g^*_1\hat{a}^{\dagger}_1\hat{b}^{\dagger}_1
+ g^*_2\hat{b}^{\dagger}_3)/\Delta $, which come from the
adiabatic elimination process in Heisenberg equations
$i\dot{\hat{b}_2}=[\hat{b}_2, \hat{H}_I]=0$ and
$i\dot{\hat{b}^{\dagger}_2}=[\hat{b}^{\dagger}_2, \hat{H}_I]=0$,
respectively. From Eq.(\ref{8}) we can obtain the following
effective  Hamiltonian
\begin{eqnarray}
\label{9}
\hat{H'}_{I}&=&2\omega'_1\hat{b}^{\dagger}_1\hat{b}_1 +\omega'_3 \hat{b}^{\dagger}_3\hat{b}_3+ (g'\hat{b}^{\dagger}_3\hat{a}_1\hat{b}_1
+ g'^*\hat{b}^{\dagger}_1\hat{a}^{\dagger}_1\hat{b}_3)\nonumber \\
& &+\lambda_1\hat{b}^{\dagger 2}_1\hat{b}^2_1
+\omega'_{1}\hat{a}^{\dagger}_1\hat{a}_1\hat{b}^{\dagger}_1\hat{b}_1,
\end{eqnarray}
which contains only atomic field operators in internal states $|1\rangle$ and $|3\rangle$ and the probe field operators. Here we have set $\hbar=1$ and introduced
\begin{equation}
\label{10}
\omega'_1=-\frac{|g_1|^2}{\Delta}, \hspace{0.3cm}
\omega'_3=-\frac{|g_2|^2}{\Delta}, \hspace{0.3cm}
g'=-\frac{g_1g^*_2}{\Delta}.
\end{equation}

The first part $\hat{a}_1\hat{b}^{\dagger}_3\hat{b}_1$ in the
third term of Eq. (\ref{9}) describes such a quantum transition
process where an atom in the lowest internal state $|1\rangle$
absorbs a photon in the probe laser and then it transits  to the
intermediate internal state $|3\rangle$. The second part
$\hat{b}^{\dagger}_1\hat{a}^{\dagger}_1\hat{b}_3$ describes the
inverse process of this process.

Then from the Hamiltonian (\ref{9}) the atomic field operators in the third internal state
can be adiabatically eliminated through the replacements:
$\hat{b}_3 = -g_1\hat{a}_1\hat{b}_1/g_2$ and $\hat{b}^{\dagger}_3 = -g^*_1\hat{a}^{\dagger}_1\hat{b}^{\dagger}_1/g^*_2$,
and the final effective Hamiltonian is given by
\begin{equation}
\label{11} \hat{H}_{eff}=2\omega'_1\hat{b}^{\dagger}_1\hat{b}_1
+4\omega'_1\hat{b}^{\dagger}_1\hat{b}_1\hat{a}^{\dagger}_1\hat{a}_1+\lambda_1
\hat{b}^{\dagger 2}_1\hat{b}^2_1.
\end{equation}

 The effective Hamiltonian (\ref{11}) is diagonal in the Fock space with the bases defined by
\begin{equation}
\label{12}
|n,m\rangle =\frac{1}{\sqrt{n!m!}}\hat{a}^{\dagger n}_1\hat{b}^{\dagger m}_1|0,0\rangle ,
\end{equation}
which are eigenstates of the effective Hamiltonian with the eigenvalues given by
the following expression
\begin{equation}
\label{13}
 E(n,m)=2\omega'_1m + 4\omega'_1nm + \lambda_1m(m-1),
\end{equation}
where $n$ and $m$ take non-negative integers.

From the effective Hamiltonian (\ref{11}) we can well understand
the theoretical mechanism to create atom-photon quantum
entanglement in present scheme. Comparing (\ref{11}) with
(\ref{8}) we can see that for the BEC in the EIT, through
adiabatically eliminating the atomic field operators at the upper
and intermediate levels the laser-atom dipole interactions are
converted as an effective collisional interaction between photons
in the probe laser and condensed atoms in the internal state
$|1\rangle$ which can be described as effective elastic collisions
between photons in the probe laser and condensed atoms in the
internal state $|1\rangle$. From Eq. (\ref{10}) we can know that
the effective scattering length to describe the atom-photon
collisions is proportional to the ratio of the square of the
dipole interaction and the two-photon detuning. One can manipulate
signs of the effective scattering length by changing signs of the
two-photon detuning. Therefore, atom-photon quantum entanglement
is produced by atom-photon and atom-atom collisional interactions
described by the second and third terms in the effective
Hamiltonian (\ref{11}), respectively, and one can control or
manipulate atom-photon entanglement by varying the dipole
interaction strength between the probe laser and the BEC and the
two-photon detuning to create desired atom-photon entangled
states.

It should be mentioned that in most textbook introductions to EIT
\cite{ari}, both the probe and coupling lasers are treated
classically, the decay of the excited state is included in the
dynamics of the internal state through a set of density-matrix
equations of the atomic system,  the decay promotes the trapping
of the atom into a dark state. Contrary to the usual treatment of
EIT, the probe laser is quantized in this paper. A limitation of
our present treatment is that we have ignored the decay rates of
various levels. However, this ignorance of the decay rates is a
good approximation for the adiabatic EIT that we study in the
present paper, since the adiabatic  EIT is insensitive to any
possible decay of the top level \cite{lm}.

\section{Atom-photon entangled states}

In this section we investigate generation of various atom-photon entangled
states when the condensate in internal states $|1\rangle$ and the probe laser are initially in non-entangled coherent states and superposition states of coherent states.

We assume that the probe laser and the condensate are initially uncorrelated and in the product coherent state $|\alpha, \beta\rangle\equiv |\alpha\rangle \otimes|\beta\rangle$. With the laser fields turned on at $t=0$, then at time $t>0$ the state of the system becomes
\begin{eqnarray}
\label{14}
|\Phi(\tau)\rangle &=& \sum^{\infty}_{n,m=0}
\exp \{-it[2\omega'_1m + 4\omega'_1nm + \lambda_1m(m-1)]\} \nonumber \\
 & &\times \exp\left[-\frac{1}{2}(|\alpha|^2+|\beta|^2)\right]\frac{\alpha^n\beta^m}{\sqrt{n!m!}} |n,m\rangle.
\end{eqnarray}

In what follows we shall see that starting with the state (\ref{14}) different atom-photon entangled states can be obtained at different times through adjusting various interaction strengths and the two-photon detuning.
In order to do this, we rewrite the state (\ref{14}) as the following simple form
\begin{equation}
\label{15}
|\Phi(\tau)\rangle =e^{-\frac{1}{2}(|\alpha|^2+|\beta|^2)} \sum^{\infty}_{n,m=0}e^{i\tau \theta_{n,m}}
\frac{\alpha^n\beta^m}{\sqrt{n!m!}}|n,m\rangle,
\end{equation}
where we have used a scaled time $\tau=\lambda_1t$ and a running frequency
\begin{equation}
\label{16}
\theta_{n,m}=(1+K)m+2Knm-m^2,
\end{equation}
where we have introduced a real effective interaction parameter $K$ defined by
\begin{equation}
\label{17}
K=\frac{2|g_1|^2}{\lambda_1\Delta},
\end{equation}
which describes an effective interaction induced by three adjustable parameters:
the dipole interaction strength $ g_1$, the inter-atomic interaction strength $\lambda_1$, and the two-photon detuning $\Delta$. The effective interaction parameter $K$ can takes positive or negative values which depends on the signs of $\lambda_1$ and $\Delta$. For an atomic BEC with a positive $s$-wave scattering length, $K$ is positive (negative) when the two-photon detuning $\Delta$ is positive (negative). From Eqs. (\ref{15}) and (\ref{16}) we can see that the time evolution characteristic of the system under our consideration is completely determined by the effective interaction parameter.

We note that the wavefunction of the system (\ref{15}) is a two-mode extension of generalized coherent states \cite{tit,bia,sto}. These generalized coherent states differ from a conventional Glauber coherent state by an extra phase factor appearing in the decomposition of such states into a superposition of Fock states. They can always be represented as a continuous sum of coherent states. And under appropriate periodic conditions, they can reduce to discrete superpositions of coherent states. It should be kept in mind that we use to create continuous-variable-type entangled states what we expect. For the two-mode case under our consideration, we can express the state (\ref{15}) as the following integral form
\begin{equation}
\label{18}
|\Phi(\tau)\rangle = \int^{2\pi}_0\frac{d\phi_1}{2\pi}
                     \int^{2\pi}_0\frac{d\phi_3}{2\pi}
 f(\phi_1, \phi_3)\left|\alpha e^{i\phi_1}, \beta e^{i\phi_3}\right\rangle,
\end{equation}
where the phase function is given by
\begin{equation}
\label{19}
 f(\phi_1, \phi_3)=\exp\left[i\left(\tau\theta_{n,m} - n\phi_1
- m\phi_3\right)\right].
\end{equation}

Eq. (\ref{19}) indicates that the state $|\Phi(\tau)\rangle $ is a continuous superposition state of two-mode product coherent states. Note that entangled states what we expect are generally superposition states of two-mode product coherent states.

From Eqs. (\ref{16}) to (\ref{19}) we can see that the values of the real parameter $K$ may seriously affect the form of the wavefunction (\ref{18}). Of particular interesting is a situation where $K$ may take values of nonzero integers. In this case, making use of (\ref{16}) from (\ref{15}) we can see that $|\Phi(\tau+2\pi)\rangle=|\Phi(\tau)\rangle$, which implies that the time evolution of the wavefunction (\ref{15}) is a periodic evolution with respect to scaled time $\tau$ with the period $2\pi$. On the other hand, suppose that the scaled time $\tau$ takes its values in the following  manner
\begin{equation}
\label{20}
\tau= \frac{M}{N}2\pi,
\end{equation}
where $M$ and $N$ are mutually prime integers, then we can find
\begin{equation}
\label{21}
\exp\left[i2\pi\frac{M}{N}\theta_{n+N, m+N}\right]
=\exp\left[i2\pi\frac{M}{N} \theta_{n, m}\right],
\end{equation}
which means that the exponential phase function in the state (\ref{19}) $\exp\left[2\pi i(M/N)\theta_{n, m}\right]$ is a periodic function with respect to both $n$ and $m$ with the same period $N$. If $\tau$ takes its values according to Eq. (\ref{20}), as a fraction of the period, then the wavefunction (\ref{18}) becomes a discrete superposition state of product coherent states which can be expressed as follows
\begin{equation}
\label{22}
\left|\Phi\left(\tau=\frac{M}{N}2\pi\right)\right\rangle = \sum^{N}_{r=1} \sum^{N}_{s=1}c_{rs}\left|\alpha e^{i\varphi_r}, \beta e^{i\varphi_s} \right\rangle,
\end{equation}
where the two running phases are defined by
\begin{equation}
\label{23}
\varphi_r=\frac{2\pi}{N}r, \hspace{0.5cm} \varphi_s=\frac{2\pi}{N}s, \hspace{0.5cm} (r, s=1,2, \cdots, N).
\end{equation}

From Eqs. (\ref{15}) and (\ref{22}) we can find the following equation to determine the coefficients $c_{rs}$,
\begin{equation}
\label{24}
\sum^{N}_{r,s=1}c_{rs}\exp\left\{\frac{2\pi i}{N}\left[nr+ms-M \theta_{n, m} \right]\right\}=1.
\end{equation}

Carrying out summations over $n$ and $m$ in both sides of the above equation from $1$ to $N$, and taking into account the normalization condition
$\sum^{N}_{r,s=1}c_{rs} c^*_{rs}=1$, we find the coefficients $c_{r s}$ to be given by
\begin{equation}
\label{25}
c_{rs}=\frac{1}{N^2}\sum^{N}_{n,m=1} \exp\left\{-\frac{2\pi i }{N}\left[nr + ms -M \theta_{n, m}\right]\right\}.
\end{equation}

It is straightforward to see that the discrete superposition state (\ref{22}) is generally an entangled coherent
state with $N^2$ independent product coherent states.  As a specific example of creating continuous-variable-type
entangled states, in what follows we discuss generation of entangled states for the case of $K=-1$.

{\it Two-state entangled states.} Assume that the probe laser and
the condensate are initially in coherent states $|\alpha\rangle$
and $|\beta\rangle$, respectively. When $K=-1$, $N=4$, and $M=1$,
from Eqs. (\ref{16}) and (\ref{25}) we find that nonzero
$c$-coefficients are
\begin{equation}
\label{26}
c_{22}=-c_{24}=ic_{42}=ic_{44}=\frac{i}{2},
\end{equation}
which results in the following unnormalized two-state entangled
state
\begin{equation}
\label{27} \left|\Phi\left(\tau=\frac{\pi}{2}\right)\right\rangle
=\frac{1}{2}[ \beta_+|\alpha\rangle\otimes |\beta\rangle_+ -
i\beta_- |-\alpha\rangle\otimes |\beta\rangle_-],
\end{equation}
where $|\beta\rangle_{\pm}$ are two normalized atomic
Schr\"odinger cat states, i.e.,  even and odd coherent states
defined by
\begin{equation}
\label{28} |\beta\rangle_{\pm}=
\frac{1}{\beta_{\pm}}(|\beta\rangle \pm
|-\beta\rangle),\hspace{0.3cm} \beta\pm=\sqrt{2(1\pm
e^{-2|\beta|^2})}.
\end{equation}

Hence, the state (\ref{27}) is an entangled state between two
photon-coherent states and two atomic Schr\"odinger cat states
where one member is photons in the probe laser the other is atoms
in the condensate.

The degree of quantum entanglement of the two-state entangled
states (\ref{27}) can be measured in terms of the concurrence
\cite{wan,hill} which is generally defined for discrete-variable
entangled states to be \cite{hill}
\begin{equation}
\label{28-01}
\mathcal{C}=|\langle\Psi|\sigma_y\otimes\sigma_y|\Psi^*\rangle|,
\end{equation}
where $|\Psi^*\rangle$ is the complex conjugate of $|\Psi\rangle$.
The concurrence equals one for a maximally entangled state.

In order to calculate the concurrence of continuous-variables-type
entangled states like (\ref{27}), we consider a general biparticle
entangled state
\begin{equation}
\label{28-2} |\Psi\rangle=\mu|\eta\rangle\otimes|\gamma\rangle +
\nu|\xi\rangle\otimes|\delta\rangle,
\end{equation}
where $|\eta\rangle$ and $|\xi\rangle$ are {\it normalized} states
of subsystem $1$ and  $|\gamma\rangle$ and $|\delta\rangle$
normalized states of subsystem $2$ with complex $\mu$ and $\nu$.
After normalization, the biparticle state $|\Psi\rangle$ can be
expressed as
\begin{equation}
\label{28-3}
|\Psi\rangle=\frac{1}{N}[\mu|\eta\rangle\otimes|\gamma\rangle +
\nu|\xi\rangle\otimes|\delta\rangle],
\end{equation}
where the normalization constant is given by
\begin{eqnarray}
\label{28-4} N^2&=&|\mu|^2+|\nu|^2 + 2 Re (\mu^*\nu p_1p^*_2),
\nonumber \\
 p_1&=&\langle\eta|\xi\rangle, \hspace{0.3cm} p_2=\langle\delta|\gamma\rangle.
\end{eqnarray}

Assume that $|\eta\rangle$ and $|\xi\rangle$ ( $|\gamma\rangle$
and $|\delta\rangle$)are linearly independent and span a $2$-dim
subspace of the Hilbert space. Then one can choose a discrete
orthogonal basis $\{|0\rangle_i,|1\rangle_i\}$ as in Ref
\cite{man}
\begin{eqnarray}
\label{28-5} |0\rangle_1&=&|\eta\rangle, \hspace{0.3cm}
|1\rangle_1=\frac{1}{\sqrt{1-|p_1|^2}}(|\xi\rangle-p_1|\eta\rangle),
\nonumber \\
|0\rangle_2&=&|\delta\rangle, \hspace{0.3cm}
|1\rangle_2=\frac{1}{\sqrt{1-|p_2|^2}}(|\gamma\rangle-p_1|\delta\rangle).
\end{eqnarray}

By using above discrete orthogonal basis, one can express the
biparticle state $|\Psi\rangle$  as the following two-qubit
entangled state
\begin{eqnarray}
\label{28-6} |\Psi\rangle&=&\frac{1}{N}\left[(\mu p_2+\nu
p_1)|0 0\rangle + \mu\sqrt{1-|p_2|^2}|0 1\rangle \right.\nonumber \\
& & \left.+\nu\sqrt{1-|p_1|^2}|1 0\rangle\right],
\end{eqnarray}
which can be written in terms of a Schmidt decomposition
\cite{man} as
\begin{equation}
\label{28-7} |\Psi\rangle=c_+|+ +\rangle + c_-|- -\rangle,
\end{equation}
where $|+ \rangle_i$ and $|- \rangle_i$ are the orthonormal
eigenvectors of the reduced density operators for the state
(\ref{28-6}) and $c_{\pm}=\sqrt{\lambda_{\pm}}$ are the square
roots of the corresponding eigenvalues given by
\begin{equation}
\label{28-8} \lambda_{\pm}=\frac{1}{2} \pm \frac{1}{2}\sqrt{1 -
\frac{4|\mu\nu|}{N^2}\sqrt{(1-|p_1|^2)(1-|p_2|^2)}}.
\end{equation}

 Making use of the Schmidt decomposition
(\ref{28-7}) and Eq. (\ref{28-8}), from Eq. (\ref{28-01}) it is
easy to find the concurrence of the entangled state (\ref{28-2})
to be \cite{run}
\begin{eqnarray}
\label{28-9} \mathcal{C}&=&2c_+c_-  \nonumber\\
                        &=&\frac{2|\mu||\nu|}{N^2}\sqrt{(1-|p_1|^2)(1-|p_2|^2)}.
\end{eqnarray}

 For the two-state entangled state (\ref{27}), from  (\ref{28-9}) we
find the corresponding concurrence
\begin{equation}
\label{29} \mathcal{C}=\sqrt{\left[1-\exp(-4|\alpha|^2)\right]
\left[1-\exp(-4|\beta|^2)\right] },
\end{equation}
which indicates that the concurrence of the two-state entangled
state (\ref{27}) is dependent of the values of the initial state
parameters of the condensate and the probe laser. From Eq.
(\ref{29}) we can also see that one can manipulate quantum
entanglement of the two-state entangled states (\ref{27}) by
varying the intensity of the probe laser and the initial number of
atoms in the condensate. In particular, in the parameter regime in
which  one can adiabatically eliminate the top level state and the
second ground internal state i.e., $(g_{1}/g_{2})^{2}<<1$, the
stronger the the intensity of the probe laser is or/and the more
the number of atoms in the condensate are, the larger the amount
of entanglement of the state (\ref{27}).

It is interesting to note that the state (\ref{27}) can also be
expressed as an entangled state of two atomic coherent states
with two photon Schr\"odinger cat states,
\begin{equation}
\label{27-1}
\left|\Phi\left(\tau=\frac{\pi}{2}\right)\right\rangle
=\frac{1}{2}[ |\alpha\rangle_-\otimes |\beta\rangle +
|\alpha\rangle_+\otimes|-\beta\rangle],
\end{equation}
where $|\alpha\rangle_{\pm}$ are two normalized photon
Schr\"odinger cat states defined by
\begin{equation}
\label{28-1} |\alpha\rangle_{\pm}= \frac{1}{\sqrt2}(|\alpha\rangle
\pm i|-\alpha\rangle).
\end{equation}

Although the states (\ref{27}) and  (\ref{27-1}) are two different
decompositions of the same state of the system under our
consideration, they have the same concurrence, which means that
their degrees of entanglement are the same.

In order to obtain atom-photon entangled coherent states, we
assume that  the probe laser is initially in a coherent photon cat
state while the condensate is initially in a coherent state.
Namely, the initial state of the atom-photon system is
\begin{equation}
\label{30}
\left|\Phi\left(\tau=0\right)\right\rangle =\frac{1}{\sqrt2}(|\alpha\rangle + i|-\alpha\rangle)\otimes|\beta\rangle.
\end{equation}

When $K=-1$, $N=4$, and $M=1$, a $\pi/2$ evolution drives the
system to the  following entangled coherent state
\begin{equation}
\label{31}
\left|\Phi\left(\tau=\frac{\pi}{2}\right)\right\rangle =\frac{1}{\sqrt2} (|\alpha \rangle\otimes| \beta \rangle
+ i|-\alpha \rangle\otimes |-\beta\rangle),
\end{equation}
where we have used Eq. (\ref{27}). The entangled state (\ref{31})
is a coherent superposition  state of two distinct pairs of
correlated coherent states of the probe laser and the atomic
condensate. It is a two-mode extension of a Yurke-Stoler state
\cite{yur}. This superposition state can be interpreted in the
spirit of Schr\"odinger's Gedanken experiment, where the different
coherent states replace the states of the cat being dead and
alive.

Similarly, when the initial state of the atom-photon system is
\begin{equation}
\label{32}
\left|\Phi\left(\tau=0\right)\right\rangle =\frac{1}{\sqrt2} (|\alpha\rangle - i|-\alpha\rangle)\otimes|\beta\rangle,
\end{equation}
if $K=-1$, $N=4$, and $M=1$,  then a $\pi/2$ evolution drives the system to the following entangled coherent state
\begin{equation}
\label{33}
\left|\Phi\left(\tau=\frac{\pi}{2}\right)\right\rangle =\frac{1}{\sqrt2} (|\alpha\rangle\otimes |-\beta \rangle
- i|-\alpha\rangle\otimes | \beta\rangle).
\end{equation}

The two two-state entangled coherent states (\ref{31}) and
(\ref{33}) have the same degree of quantum entanglement indicated
by the concurrence given by
\begin{equation}
\label{34}
\mathcal{C}=\sqrt{\left[1-\exp(-4|\alpha|^2)\right] \left[1-\exp(-4|\beta|^2)\right] },
\end{equation}
which implies that the amount of entanglement of the two two-state
entangled coherent states (\ref{31}) and (\ref{33}) increase with
$\alpha$ and $\beta$ in the regime where the conditions of the
adiabatic elimination of the related levels are satisfied.

{\it Three-state entangled states.} In order to obtain photon-atom
three-state entangled states, we consider a situation where the
probe laser and the condensate are initially in coherent states
$|\alpha\rangle$ and $|\beta\rangle$, respectively, while $K=-1$,
$N=3$, and $M=1$. In this case from Eqs. (\ref{15}) and (\ref{25})
we find that nonzero $c$-coefficients are given by
\begin{eqnarray}
\label{35}
c_{11}&=&c_{22}=c^*_{13}=c^*_{23}=-\frac{1}{3}e^{-i\frac{\pi}{3}}, \nonumber \\
c_{12}&=&c_{21}=c_{31}=c_{32}=c_{33}=\frac{1}{3},
\end{eqnarray}
which results in  the following three-state entangled  state
\begin{eqnarray}
\label{36}
\left|\Phi\left(\tau=\frac{2\pi}{3}\right)\right\rangle &=&
\frac{1}{3} \left[ |\alpha\rangle_1\otimes|\beta\rangle_1
+ |\alpha\rangle_2\otimes|\beta\rangle_2 \right.\nonumber \\
& & \left.+ e^{-i\frac{\pi}{3}}|\alpha\rangle_3\otimes|\beta\rangle_3\right],
\end{eqnarray}
where we have discarded the common phase factor $\exp(i\pi/3)$ on
the right-hand  side of above equation and  three photon-coherent
states and three atomic MQS states are defined by
\begin{eqnarray}
\label{37}
|\alpha\rangle_k&=&|(-1)^k \alpha e^{-i\frac{k\pi}{3}}\rangle, \hspace{0.5cm} (k=1,2,3)\nonumber \\
|\beta\rangle_1&=&e^{-i\frac{\pi}{3}}|-\beta e^{i\frac{\pi}{3}}\rangle
+ e^{i\frac{\pi}{3}}|-\beta e^{-i\frac{\pi}{3}}\rangle - |\beta\rangle, \nonumber\\
|\beta\rangle_2&=&e^{i\frac{\pi}{3}}|-\beta e^{i\frac{\pi}{3}}\rangle
+ e^{-i\frac{\pi}{3}}|-\beta e^{-i\frac{\pi}{3}}\rangle - |\beta\rangle, \nonumber\\
|\beta\rangle_3&=&|-\beta e^{i\frac{\pi}{3}}\rangle
+ |-\beta e^{i\frac{\pi}{3}}\rangle + |\beta\rangle.
\end{eqnarray}
Hence, the state (\ref{36}) is an entangled state between three
photon-coherent  states and three atomic  MQS states.

{\it Four-state entangled states.} Suppose that the probe laser
and the condensate  are initially in coherent states
$|\alpha\rangle$ and $|\beta\rangle$, respectively. We consider a
$\pi/4$ evolution of the system, $K=-1$, $N=8$, and $M=1$, From
Eqs. (\ref{15}) and (\ref{25}) we find that nonzero
$c$-coefficients are given by
\begin{eqnarray}
\label{38}
c_{22}&=&-c^*_{24}=-c_{26}=c^*_{28}=-c_{62} \nonumber \\
&=&-c^*_{64}=c_{66}=c^*_{68}=\frac{1}{4}\exp{\left(i\frac{\pi}{4}\right)},
 \nonumber \\
c_{42}&=&-c_{44}=c_{46}=-c_{48}=c_{82}\nonumber \\
&=&c_{84}=c_{86}=c_{88}=\frac{1}{4},
\end{eqnarray}
which result in  the following four-state entangled  state
\begin{eqnarray}
\label{39}
\left|\Phi\left(\tau=\frac{\pi}{4}\right)\right\rangle &=&
\frac{1}{4} \left[ e^{i\frac{\pi}{4}}|i\alpha\rangle_-\otimes|i\beta\rangle_-
+ e^{-i\frac{\pi}{4}}|i\alpha\rangle_+\otimes|\beta\rangle_- \right.\nonumber \\
& & \left.+ |\alpha\rangle_+\otimes|i\beta\rangle_+
+ |\alpha\rangle_-\otimes|\beta\rangle_+ \right],
\end{eqnarray}
where $|\gamma\rangle_{\pm}=|\gamma\rangle \pm |-\gamma\rangle$
with $\gamma=\alpha$, $\beta$,  $ i\alpha$ and  $i\beta$ are
unnormalized Schr\"odinger cat states. Hence, the state (\ref{39})
is an entangled state between two pairs of photon Schr\"odinger
cat states and two pairs of  atom Schr\"odinger cat states.

It is worthwhile to mention that besides their importance on
quantum entanglement  for above two-state, three-state, and
four-state entangled states, we note that all of them are
superposition states of a macroscopic number of atoms with a
macroscopic number of photons. In this sense, we have also given a
scheme for generating atom-photon Schr\"{o}dinger cat states. The
problem of creating a macroscopic superposition of atoms and
photons raises an interest itself, because rather than two states
of a given object, the atom-photon system is a seemingly
impossible macroscopic superposition of two different objects.
This is beyond the usual macroscopic superposition of two states
of a given object, the atom-photon system actually leads to a more
counterintuitive situation since atoms and photons are different
objects.

\section{Concluding Remarks}
We have presented a scheme for the generation of atom-photon
entangled states of  continuous-variable-type pure states in the
atomic BEC which exploits quantum interference, or EIT, in
three-level atoms. We have shown how to generate multi-state
entangled coherent states when the atom-photon system is initially
in an uncorrelated product coherent state. As examples, the
generation of two-state, three-state, and four-state entangled
states has been investigated explicitly. We have created not only
atom-photon entangled coherent states, but also entangled states
between photon (atom)-coherent states and atom (photon)-cat
states, and between photon-cat and atom-cat states. All of theses
entangled states are MQS states of the atom-photon system. They
may be regarded as an extension of the usual Schr\"{o}dinger cat
states of one given object to two different objects (atoms and
photons).

In the proposed scheme we use the atomic BEC with three internal
states which are coupled with a weak quantized probe laser and a
strong classical coupling laser. When the BEC is in the EIT, since
there is no atomic population at the upper and intermediate
levels, the atomic field operators at these levels may be
adiabatically eliminated. The system then becomes an effective
two-mode system in which one member is the probe laser (photons),
the other is the condensate (atoms) in the internal state
$|1\rangle$. In this process the (probe) laser-atom dipole
interaction is converted as an atom-photon effective collisional
interaction. The strength of the atom-photon collisional
interaction is determined by the dipole interaction and the
two-photon detuning. Atom-photon entanglement is produced by
inter-atomic and {\it atom-photon effective collisional
interactions}. This mechanism to create quantum entanglement is
different with that in our previous analysis of generating
atom-atom continuous-variable-type entangled states in a
three-level $\Lambda$-shaped BEC system \cite{kuan}. Atom-atom
entanglement in Ref. \cite{kuan} is produced by inter-atomic
collisional interaction and {\it an inter-atomic effective
tunnelling interaction} generating by adiabatically eliminating
the atomic field operators at the upper level, and no EIT is
required. In present scheme the technical requirements involve two
lasers with a well-controlled frequency difference, a controllable
dipole interaction between the probe laser and BEC, and a
manipulable scattering length for atoms in the internal state
$|1\rangle$. These are met by most existing EIT and BEC
experiments. We hope that the proposed method for generating
atom-photon continuous-variable-type entangled states can find its
applications in quantum information processing and studies of
fundamental problems of quantum mechanics. It is possible to apply
the approach to create entangled states in the present paper to
study quantum entanglement between two distant condensates and
quantum teleportation of quantum states of BEC's, however, it is
outside the scope of the present paper and will be discussed
elsewhere.

\acknowledgments

L.M. Kuang would like to thank Professor Yong-Shi Wu and Dr.
Guang-Hong Chen  for useful discussions. This work is supported by
the National Fundamental Research Program Grant No. 2001CB309310,
the National Natural Science Foundation Grant Nos. 90203018 and
10075018, the State Education Ministry of China, the Educational
Committee of Hunan Province, and the Innovation Funds from Chinese
Academy of Sciences via the Institute of Theoretical Physics,
Academia, Sinica.

\end{document}